\documentclass[twocolumn,prb]{revtex4}
\usepackage{graphicx}    
\usepackage{dcolumn}     
\usepackage{bm}          
\newcommand{\Vec}[1]{\mbox{\boldmath$#1$}}
\begin{document}
\preprint{qmctriplet}

\title{Quantum Monte Carlo study of the pairing symmetry competition in 
the Hubbard model
}
\author{
Kazuhiko Kuroki
}
\affiliation{
Department of Applied Physics and
Chemistry, The University of Electro-Communications,
Chofu, Tokyo 182-8585, Japan}

\author{Yukio Tanaka}
\affiliation{Department of Applied Physics,
Nagoya University, Nagoya, 464-8603, Japan}

\author{
Takashi Kimura and Ryotaro Arita}
\affiliation{Department of Physics,
University of Tokyo, Hongo 7-3-1, Tokyo 113-0033, Japan}

\date{\today}
\begin{abstract}
To shed light into the pairing mechanism of 
possible spin-triplet superconductors 
(TMTSF)$_2$X and Sr$_2$RuO$_4$,
we study the competition among various spin singlet and 
triplet pairing channels in the Hubbard model by calculating the 
pairing interaction vertex using the ground state 
quantum Monte Carlo technique.
We model (TMTSF)$_2$X by 
a quarter-filled quasi-one dimensional (quasi-1D) Hubbard model,
and the $\gamma$ band of Sr$_2$RuO$_4$ 
by a two dimensional (2D) Hubbard model with a band filling of $\sim 4/3$.
For the quasi-1D system, we find that triplet $f$-wave pairing 
not only dominates over triplet $p$-wave in agreement with 
the spin fluctuation theory, but also looks unexpectedly 
competitive against $d$-wave. 
For the 2D system, although the results suggest presence of 
attractive interaction in the triplet pairing channels, the $d$-wave
pairing interaction is found to be larger than those of the triplet 
channels.
\end{abstract}
\maketitle
\section{Introduction}
Possible occurrence of spin-triplet pairing in superconductors 
such as a ruthenate Sr$_2$RuO$_4$\cite{Maeno,Luke,Ishida,Duffy} 
and organic materials (TMTSF)$_2$X (X=ClO$_4$, PF$_6$ ...)\cite{Lee,Lee2} 
has attracted much attention lately. A probable mechanism 
of spin-triplet pairing is due to ferromagnetic spin fluctuations, as in 
$^3$He, as proposed in ref.\onlinecite{RS} for Sr$_2$RuO$_4$.
In both of these materials, however, 
the mechanism of triplet pairing has been 
controversial because the spin fluctuation has
turned out to be more like antiferromagnetic than ferromagnetic. 
Namely in (TMTSF)$_2$X, the nesting of the quasi-1D Fermi surface 
give rise to a $2k_F$ spin density wave phase lying right next to 
the superconducting phase.
In Sr$_2$RuO$_4$, spin fluctuations 
with a wave number $(2\pi/3,2\pi/3)$ arises due to 
the nesting of the quasi-1D $\alpha$ and $\beta$ bands.\cite{Sidis}

For (TMTSF)$_2$X, the possibility of spin triplet $p$-wave pairing has been 
considered from the early days,\cite{Emery,Lebed,Fukuyama}
but if we consider the $2k_F$ spin fluctuations to be the origin of the 
pairing interaction,   
spin singlet $d$-wave-like pairing is more likely to occur   
as proposed by several authors.\cite{Shimahara,KiKo,KA,Nomura}
Recently, two of the present authors have proposed,
based on a phenomelogical spin-charge fluctuation theory, that 
a combination of quasi-one-dimensionality, the coexistence of 
$2k_F$ spin and charge fluctuations, and anisotropy in the spin fluctuations 
may lead to spin-triplet
$f$-wave pairing, dominating over $d$- and $p$-wave pairings.\cite{KAA}
A similar proposal has been made by Fuseya et al.\cite{Fuseya1}
In fact, the coexistence of $2k_F$ spin and charge density wave 
has been reported experimentally.\cite{PR,Kagoshima}
However, a recent renormalization group study by Fuseya {\it et al.}
\cite{Fuseya2} 
has suggested that $p$-wave pairing dominates over $f$-wave pairing
in quasi-1D systems.

For Sr$_2$RuO$_4$ on the other hand, 
there has been a debate as to which one of the three bands plays the 
main role in the occurrence of superconductivity, namely 
the 2D $\gamma$ bands or the quasi-1D $\alpha$-$\beta$ bands.
Several microscopic theories have 
focused on the antiferromagnetic spin 
fluctuations due to the 
$\alpha$-$\beta$ band Fermi surface nesting.\cite{KO,SK,KOAA,Takimoto,Eremin}
In refs.\onlinecite{KO}-\onlinecite{KOAA} in particular, 
the experimentally observed 
anisotropy \cite{Mukuda} in the spin fluctuations 
is the key for spin-triplet $p$-wave pairing to take place.

On the other hand, Nomura and Yamada focused mainly on the 
2D $\gamma$ band.\cite{NY}
Applying third order perturbation theory to the 2D Hubbard model 
with appropriate band structure and band filling, they showed 
that triplet pairing superconductivity takes place, 
dominating over singlet pairing. A multiband calculation 
has recently been performed along this line.\cite{Yanase}
Also, a renormalization group study has been performed for the 
Hubbard model having nearest neighbor hopping, where spin triplet 
pairing is shown to be the leading instability near but 
away from the van Hove filling.\cite{Honerkamp}
By contrast, if we apply random phase approximation (RPA) type theories 
like the fluctuation exchange method (FLEX)\cite{Bickers} 
to the 2D Hubbard model, 
triplet pairing not only does not occur in 
a realistic temperature regime,\cite{AKA,ML} 
but also does not even dominate over singlet pairing 
at the filling corresponding to the $\gamma$ band of Sr$_2$RuO$_4$ 
(see section\ref{VD}).

The purpose of the present study is to 
shed light into these controversies. 
We model (TMTSF)$_2$X and Sr$_2$RuO$_4$ 
by quasi-1D and 2D Hubbard models, respectively.
Using the ground state quantum Monte Carlo technique, we calculate
the pairing interaction vertices for possible pairing symmetries.

\begin{figure}[h]
\begin{center}
\scalebox{0.8}{
\includegraphics[width=10cm,clip]{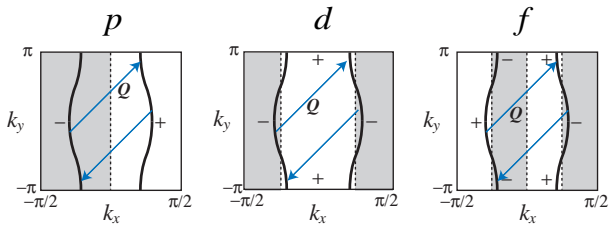}}
\caption{
$p$, $d$, and $f$-wave
gap functions along with the Fermi surface of the quasi 1D system 
and the nesting vector \Vec{Q}. The arrows denote a pair scattering
process having a momentum transfer \Vec{Q}.
\label{fig1}}
\end{center}
\end{figure}

\begin{figure}
\begin{center}
\scalebox{0.7}{
\includegraphics[width=10cm,clip]{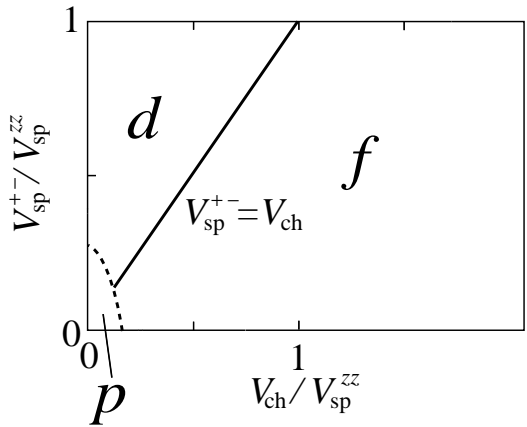}}
\caption{
A generic phase diagram for quasi-1D systems.
\label{fig2}}
\end{center}
\end{figure}

\section{Spin-charge fluctuation theory}
\label{spincharge}
Before going into the QMC calculation, 
here we  summarize the spin-charge fluctuation
theory of quasi-1D systems given in ref.\onlinecite{KAA}. 
The pairing interactions mediated by spin and charge fluctuations 
are generally given as 
\begin{eqnarray}
V_s(\Vec{q})=\frac{1}{2}V^{zz}_{\rm sp}(\Vec{q})+V^{+-}_{\rm sp}(\Vec{q})
-\frac{1}{2}V_{\rm ch}(\Vec{q})
\label{1}
\end{eqnarray}
for singlet pairing, 
\begin{eqnarray}
V_{t\perp}(\Vec{q})=
-\frac{1}{2}V^{zz}_{\rm sp}(\Vec{q})-\frac{1}{2}V_{\rm ch}(\Vec{q})
\label{2}
\end{eqnarray}
for triplet pairing with total $S_z=\pm 1$ 
($d$-vector $\Vec{d}\perp\Vec{z}$), and 
\begin{eqnarray}
V_{t\parallel}(\Vec{q})=
\frac{1}{2}V^{zz}_{\rm sp}(\Vec{q})-V^{+-}_{\rm sp}(\Vec{q})
-\frac{1}{2}V_{\rm ch}(\Vec{q})
\label{3}
\end{eqnarray}
for triplet pairing with $S_z=0$ ($\Vec{d}\parallel\Vec{z}$).
Here $V_{\rm sp}^{zz}$, $V_{\rm sp}^{+-}$, and $V_{\rm ch}$ (all positive) are
contributions from the longitudinal spin, transverse spin, and 
charge fluctuations, respectively.
We assume that the fluctuations are enhanced at $\Vec{q=Q}$,
where $\Vec{Q}$ is the nesting vector of the Fermi surface.
In the case of (TMTSF)$_2$X, $\Vec{Q}$ bridges the two disconnected 
pieces of the quasi 1D Fermi surface (see Fig.\ref{fig1}).
The pairing symmetry is determined so as to make the 
quantity 
\begin{equation}
V_{\rm eff}=
-\frac{\sum_{\Vec{k,k'}\in {\rm Fermi\: surface}} 
V_\alpha(\Vec{k-k'})\phi(\Vec{k})\phi(\Vec{k'})}
{\sum_{\Vec{k}\in {\rm Fermi\: surface}}\phi^2(\Vec{k})},
\label{4}
\end{equation}
positive and large, where $V_\alpha(\Vec{k-k'})$ is one of the pairing 
interactions (\ref{1})-(\ref{3}) and $\phi(\Vec{k})$ is the gap function. 
When $V_\alpha(\Vec{q})$ peaks around 
the nesting vector $\Vec{q}=\Vec{Q}$, the contribution around 
$\Vec{k-k'=Q}$ becomes dominant in the summation of eqn.(\ref{4}), 
so that the sign of 
the quantity $V_\alpha(\Vec{Q})\phi(\Vec{k})\phi(\Vec{k+Q})$ 
has to be negative in order to have superconductivity.

Let us now consider $p$, $d$, $f$-wave gap functions shown 
in Fig.\ref{fig1} along with the Fermi surface. 
Although the $d$- ($f$-) wave gap is not $d$ ($f$) in 
the strict sense of the word, here we use this terminology   
in the sense that the gap changes sign as $+-+-$\ \ \  ($+-+-+-$)
{\it along the Fermi surface}.
Since the $d$-wave gap function satisfies  
$\phi(\Vec{k})\phi(\Vec{k+Q})<0$, a positive 
pairing interaction $V_s(\Vec{Q})$ is necessary for superconductivity to 
occur in this channel. Similarly $V_{t\perp}(\Vec{Q})<0$ or 
$V_{t\parallel}(\Vec{Q})<0$ is necessary for $f$-wave
pairing. If we assume $V_{sp}^{zz}\geq V_{sp}^{+-}$, 
$|V_{t\perp}(\Vec{Q})|>|V_{t\parallel}(\Vec{Q})|$, so that $f$-wave
is more likely to occur in the $d\perp z$ channel. 
Since the number of nodes of the 
gap {\it intersecting the Fermi surface} is the same 
between $d$ and $f$ due to the quasi-one-dimensionality, 
the competition between the two is 
almost solely dominated by the absolute value of the 
pairing interaction. Thus the boundary between $d$ and $f$-waves
is roughly given by $V_{t\perp}(\Vec{q})=-V_s(\Vec{q})$, 
which corresponds to the line $V_{\rm ch}=V_{\rm sp}^{+-}$. In particular,
when the system has spin rotational symmetry, the two symmetries
become nearly degenerate when spin and charge fluctuations 
have about the same strength.\cite{Takimoto}

On the other hand, the $p$-wave gap function satisfies 
$\phi(\Vec{k})\phi(\Vec{k+Q})<0$, 
so that a positive triplet pairing interaction 
is necessary. This can be the case only for $\Vec{d}\parallel z$  and 
only when $V_{\rm ch}/V_{\rm sp}^{zz}$ and 
$V_{\rm sp}^{+-}/V_{\rm sp}^{zz}$ are both small.\cite{KO,SK,KOAA} 
Since the $p$-wave gap function is nodeless on the Fermi 
surface as opposed to the case of $d$- and $f$-waves, 
this will be the dominant pairing if the 
pairing interaction is positive and strong enough.

From the above phenomelogical argument, we obtain a 
generic phase diagram given in Fig.\ref{fig2}.

\section{Formulation of the QMC study}

\begin{figure}
\begin{center}
\scalebox{0.7}{
\includegraphics[width=10cm,clip]{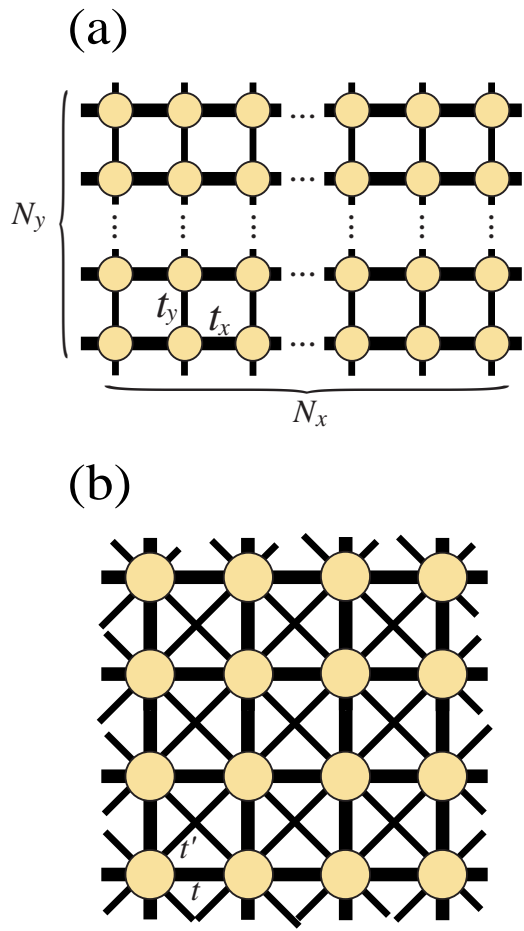}}
\caption{
The on-site Hubbard model for (a) the quasi-1D system, and (b)
the 2D system.
\label{fig3}}
\end{center}
\end{figure}

\begin{figure}
\begin{center}
\scalebox{0.8}{
\includegraphics[width=10cm,clip]{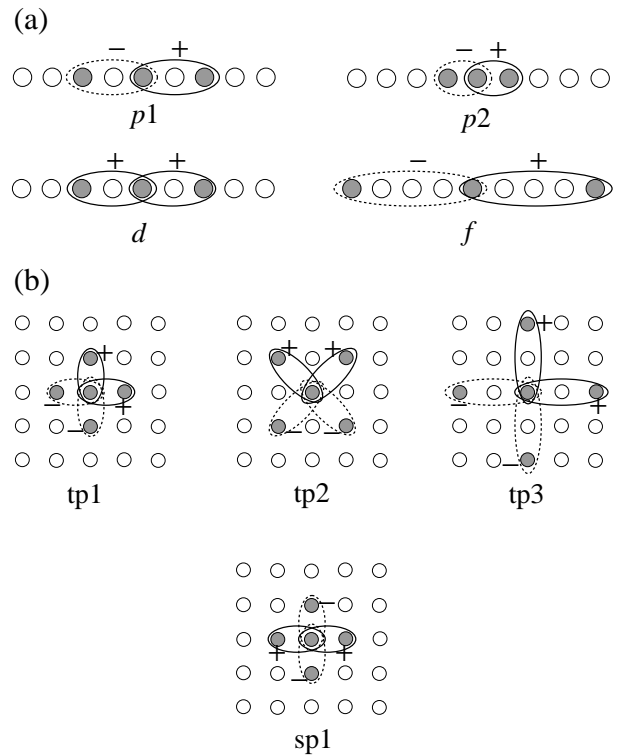}}
\caption{
The pairing channels considered for (a) the quasi-1D system, and (b)
the 2D system. The signs denote $g_\alpha(\Vec{\delta})$, 
where $\Vec{\delta}$ is measured from the center site.
\label{fig4}}
\end{center}
\end{figure}

In this section, we formulate the ground state QMC study.

\subsection{On-site Hubbard model}
As a model for (TMTSF)$_2$X, we consider the Hubbard model
\[
H=-\sum_{\langle i,j\rangle} t_{ij}(c^\dagger_{i\sigma}c_{j\sigma}+
c^\dagger_{j\sigma}c_{i\sigma})+U\sum_i n_{i\uparrow}n_{i\downarrow}
\]
on an $N_x\times N_y$ quasi-1D lattice shown in Fig.\ref{fig3}(a). 
We set $t_x\ll t_y$, where $t_x$ (taken as unity throughout the study) 
and $t_y$ are hopping
integrals in the $x$ and $y$ directions, respectively. 
Since the energy band is much more dispersive in the $x$ direction
than in the $y$ direction, we take more sites in the former 
than in the latter $(N_x\gg N_y)$.
As for the electron-electron interaction, 
we assume only the on-site repulsion $U$ for simplicity, 
and refer to the model as the on-site Hubbard model. 
The reason for this is to make the QMC calculation feasible.
Note that, $U$ enhances only the spin fluctuation
($V_{\rm sp}^{zz}=V_{\rm sp}^{+-}\gg V_{\rm ch}$), so that $d$-wave pairing 
is expected from the spin fluctuation theory\cite{Shimahara,KiKo} 
as seen from the phase diagram.
We take $U=2$ throughout the study.
The band filling $n=$(number of electrons $N_e$ /number of sites $N$) 
is taken close to quarter filling, i.e., $n\simeq 0.5$.

As for Sr$_2$RuO$_4$, we focus on the $\gamma$ band and model it by 
the 2D single band Hubbard model having nearest and next nearest neighbor 
hoppings, $t$ and $t'$ respectively, as shown in Fig.\ref{fig3}(b).
$t'$ has been estimated to be 0.3-0.4 in units of $|t|$.\cite{MS,LL,Ng} 
Here we take $t'=0.4$.
The band filling is taken to be close to $n=4/3$.

\subsection{Calculation conditions}
We adopt the auxiliary field quantum Monte Carlo method for the ground state.
\cite{SuKo,Sorella,White,IH}
In this formalism, physical quantities are evaluated using 
the ground state wave function obtained as 
\[
\exp(-\tau H)|\phi_T\rangle,
\]
where $H$ is the Hamiltonian, $\tau$ is a sufficiently large number,
and $|\phi_T\rangle$ is a trial state.
$\tau$ is sliced into a product of many $\Delta\tau$ using Trotter-Suzuki 
decomposition.\cite{Suzuki} 
In the actual calculation, we have taken $\beta\sim 10$, and 
$\Delta\tau\leq 0.1$, which turns out to be sufficient for a relatively 
small value of $U=2$. The correlation functions are evaluated as the 
average of the calculation results of about $10$ independent Monte Carlo runs, 
each of which consists of few thousand MC sweeps.
The error bars are estimated as the variation of the independent runs.

\subsection{Pairing interaction vertex}
 We calculate the pairing interaction 
vertex for possible pairing channels,\cite{White} which is defined as 
\begin{eqnarray*}
\Delta P_\alpha(i,j)&=&
\sum_{\delta,\delta'}
g_\alpha(\delta)g_\alpha(\delta')\times\nonumber\\
&&\left\{\left[\langle c_{i\uparrow}c_{i+\delta\downarrow}
c^\dagger_{j+\delta'\downarrow}c^\dagger_{j\uparrow}\rangle
-
\langle c_{i\uparrow}c^\dagger_{j\uparrow}\rangle
\langle c_{i+\delta\downarrow}c^\dagger_{j+\delta'\downarrow}\rangle\right]
\right. \\
&\pm&
\left[\langle c_{i\uparrow}c_{i+\delta\downarrow}
c^\dagger_{j\downarrow}c^\dagger_{j+\delta'\uparrow}\rangle
-
\langle c_{i\uparrow}c^\dagger_{j+\delta'\uparrow}\rangle
\langle c_{i+\delta\downarrow}c^\dagger_{j\downarrow}\rangle\right]\\
&\pm&
\left[\langle c_{i\downarrow}c_{i+\delta\uparrow}
c^\dagger_{j\uparrow}c^\dagger_{j+\delta'\downarrow}\rangle
-
\langle c_{i\downarrow}c^\dagger_{j+\delta'\downarrow}\rangle
\langle c_{i+\delta\uparrow}c^\dagger_{j\uparrow}\rangle\right]\\
&+&\left.\left[\langle c_{i\downarrow}c_{i+\delta\uparrow}
c^\dagger_{j+\delta'\uparrow}c^\dagger_{j\downarrow}\rangle
-
\langle c_{i\downarrow}c^\dagger_{j\downarrow}\rangle
\langle c_{i+\delta\uparrow}c^\dagger_{j+\delta'\uparrow}\rangle\right]
\right\}
\end{eqnarray*}
Here, the signs$\pm$ takes $+$ for 
singlet pairing, and $-$ for triplet.
$g_\alpha(\delta)$ gives sign
that depends on the pairing channel $\alpha$ as given below.
A positive interaction vertex in a certain pairing channel implies that
there is an effective attractive pairing interaction in that channel 
between the {\it renormalized} quasi-particles.

For the quasi-1D system, 
we consider the pairing channels 
 $p1$, $p2$, $d$ and $f$ as shown in Fig.\ref{fig4}(a) along with the signs
 $g_\alpha(\delta)$. It can be easily shown that the 
corresponding gap functions in momentum space are proportional to 
$\sin(2k_x)$, $\sin(k_x)$, $\cos(2k_x)$, and $\sin(4k_x)$,
respectively.
For the 2D system, 
three triplet pairings tp1, tp2, tp3 
and a singlet pairing sp1 are considered as shown in Fig.\ref{fig4}(b).
The corresponding gap functions in momentum space are  proportional to 
$\sin(k_x)$ (or equivalently $\sin(k_y)$), 
$\sin(k_x+k_y)$ (or $\sin(k_x-k_y)$), $\sin(2k_x)$ (or $\sin(2k_y)$), and 
$\cos(k_x)-\cos(k_y)$, respectively. sp1 is the so-called 
$d_{x^2-y^2}$ pairing. 

Let us denote the $i$-th site as $i=(i_x,i_y)$. For the quasi-1D system,
we calculate 
\[
\Delta P_\alpha(r)=\sum_{i_y}\Delta P_\alpha(i_x,i_y,j_x,i_y)
\]
(pairing correlation within each chain) as a function 
of distance in the $x$ direction $r=|i_x-j_x|$. For the 2D system,
we calculate 
\[
\Delta P_\alpha(r)=\sum_{i_x+i_y=r}\Delta P_\alpha(i_x,i_y,j_x,j_y)
\]
 as a function of the ``distance'' $r$.
In both systems, we focus on the long range part $\Delta P_\alpha(r)$
because superconductivity is characterized by the long ranged behavior of 
the pairing correlation function.

\subsection{Open shell and closed shell fillings}
Since the energy scale of superconductivity in repulsively 
interacting systems is small, pairing correlations are known to be 
strongly affected by the energy level spacing around the Fermi level 
in finite size systems.\cite{KTA,TKA,KA2} Namely, since superconductivity 
occurs due to pair scattering between states near the Fermi level,
the enhancement of the pairing correlation 
can be wiped out when the energies of the highest occupied levels and 
the lowest unoccupied levels are well separated.

In order to cope with this problem, 
here we consider the following two situations. (i) Open shell fillings : 
the number of electrons are chosen so that the $k$-points at the 
Fermi level is half-filled for the $U=0$ ground state. 
The highest occupied and the lowest unoccupied states 
have the same energy in this case. 
We tune the value of the hopping integrals ($t_y$ for the quasi-1D system 
and $t'$ for the 2D system)
appropriately so as to distribute the $k$-points having the same energy 
as uniformly as possible along the Fermi surface. 

In the case of open shell fillings, 
the trial state is chosen as follows. 
The $U=0$ single particle states 
($\exp(i\Vec{k}\cdot\Vec{r})$ )
below the Fermi energy is completely filled with electrons.
At the Fermi energy, the states of the form 
$\exp(i\Vec{k}\cdot\Vec{r})+\exp(-i\Vec{k}\cdot\Vec{r})$
are filled only by up spin electrons, while the states of the form 
$\exp(i\Vec{k}\cdot\Vec{r})-\exp(-i\Vec{k}\cdot\Vec{r})$
are filled only by down spin electrons.
We will use gray circles to denote the half-filled $k$-points 
in the trial state, as shown 
in the inset of Figs.\ref{fig5}, \ref{fig7}, \ref{fig8}, and \ref{fig10}(a).

(ii) Closed shell fillings : 
the number of electrons is 
chosen so that the Fermi energy for the 
$U=0$ ground state comes in between slightly separated energy levels. 
For the quasi-1D system, $t_y$ is tuned so that the energy difference 
between the highest occupied and the lowest unoccupied levels is  
within $0.01t_x$ for $U=0$. As for the 2D system, $t_y$ is taken to be 
slightly different from $t_x$ thereby lifting
the degeneracy, by less than $0.01t_x$, between the highest occupied levels 
$\Vec{k}=(k_1,k_2)$ and the lowest unoccupied levels $\Vec{k}=(k_2,k_1)$.

In this case we take the $U=0$ ground state, which is a state with 
total spin $S_{\rm tot}=0$, as the QMC trial state.
Closed (open) circles will be used to denote the occupied 
(unoccupied) $k$-points that lie within $0.01t_x$ to the Fermi energy 
in the trial state, as shown 
in the inset of Figs.\ref{fig6}, \ref{fig9}, and \ref{fig10}(b).

For open shell fillings, triplet pairing is expected to 
be favored because the electrons at the 
half-filled Fermi level tends to be in a high spin configuration.
For closed shell fillings on the other hand, 
spin singlet pairing should be favored 
since we restrict ourselves to $S_{\rm tot}=0$ states 
by choosing the $U=0$ ground state as the trial state.
This problem does make the study of singlet-triplet 
competition less conclusive,
but our results still give significant 
information concerning the $d$- vs. $f$-wave competition in the 
on-site Hubbard model as we shall see below.
As for the competition {\it within} triplet pairings 
($p$ vs. $f$) on the other hand, we can safely give a robust conclusion.

\section{Results and Discussions for the quasi 1D system}

\subsection{Open shell filling}
\label{4a}
\begin{figure}
\begin{center}
\scalebox{0.8}{
\includegraphics[width=10cm,clip]{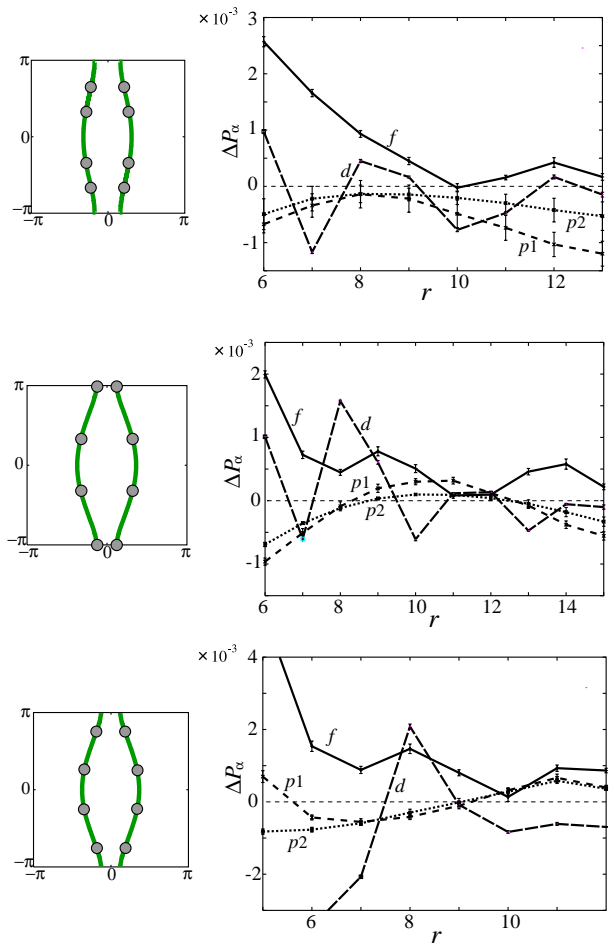}}
\caption{
Pairing interaction vertex $\Delta P_\alpha$ plotted against 
distance $r$ for the quasi-1D system with open shell fillings. 
Parameter values are taken as ($N$, $N_e$, $n$, $t_y$)= (a) ($28\times 6$, 
84, 0.50, 0.158), (b) ($32\times 6$, 96, 0.50, 0.246) ,
(c) ($24\times8$, 100,0.52, 0.259). The inset shows the half-filled 
$k$-points just at the Fermi energy in the trial state of QMC, 
along with the Fermi surface of the 
corresponding infinite size systems.
\label{fig5}}
\end{center}
\end{figure}

We now show the results for the quasi 1D system. 
We first present results for open shell fillings.
The pairing interaction vertices $\Delta P_s$ are 
shown for three different sets of parameters in Fig.\ref{fig5}.
In all cases, $\Delta P_f$ is found to be positive at 
large distances, while $\Delta P_{p1}$ and $\Delta P_{p2}$ 
are negative at most of the distances. This means that the 
spin-triplet pairing interaction is attractive in the $f$-wave channel and 
repulsive in the $p$-wave channels.
This result is 
in contrast with the recent renormalization group study\cite{Fuseya2},
and is at least qualitatively 
consistent with the spin-charge fluctuation theory described in 
section\ref{spincharge},
where $V_{\rm sp}^{zz}=V_{\rm sp}^{+-}\gg
V_{\rm ch}$ and thus $V_{t\parallel}(\Vec{Q})=V_{t\perp}(\Vec{Q})<0$ 
hold for the on-site Hubbard model.

The interaction vertex for $d$-wave pairing $\Delta P_d$
takes both positive and negative values, and we find 
$\Delta P_f>\Delta P_d$ at most of the large distances.\cite{commentd-f}
This is somewhat surprising because when the system has spin rotational 
symmetry, $f$-wave can be competitive against 
$d$-wave only when $2k_F$ spin and $2k_F$ charge fluctuations have about the 
same strength as far as the spin-charge fluctuation theory is 
concerned,\cite{KAA} which is not the case for 
the on-site Hubbard model.\cite{Hirsch} 

As mentioned before, open shell filling is expected to 
favor spin triplet pairing, so this result may be regarded as an 
artifact due to the open shell filling. However, as we shall see 
later, open shell filling does not necessarily result in 
$\Delta P_{\rm triplet} > \Delta P_{\rm singlet}$. 
Another possible reason for $f$ dominating over $d$ 
is that effects beyond RPA-type theories may be important in quasi-1D systems.
This is highly probable because Fermi liquid picture is known to 
be invalid at least in purely 1D systems.\cite{Solyom,EmeryRev}

\subsection{Closed shell filling}
\begin{figure}[htb]
\begin{center}
\scalebox{0.8}{
\includegraphics[width=10cm,clip]{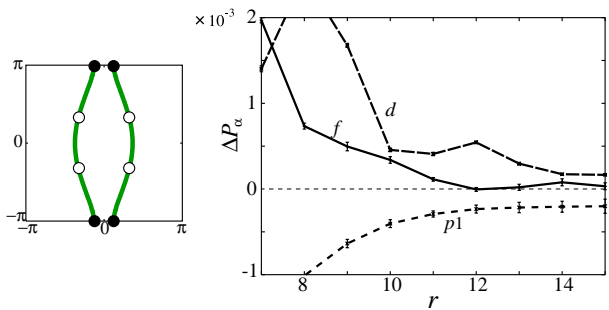}}
\caption{
A plot similar to Fig.\protect\ref{fig5} but for a closed shell
Parameter values are taken as ($N$, $N_e$, $n$, $t_y$)=($32\times 6$, 
98, 0.51, 0.240). The inset shows the 
$k$-points that lie within $0.01t_x$ to the Fermi energy 
in the trial state of the QMC calculation,
along with the Fermi surface of the 
corresponding infinite size systems.
\label{fig6}}
\end{center}
\end{figure}

Calculation results for a closed shell filling is shown in 
Fig.\ref{fig6}. In this case, we have positive $\Delta P_d$, 
smaller but still positive (or non-negative) 
$\Delta P_f$ for all distances, and clearly negative $\Delta P_p$. 
Thus, we may conclude that $\Delta P_f >\Delta P_p$ holds 
for both open and closed shell fillings. 
On the other hand, although the result of $\Delta P_d > \Delta P_f$ 
apparently seems to be consistent with the spin fluctuation theory,
this result may not be taken directly is since closed shell fillings 
selectively favors spin singlet pairing as mentioned above.
We shall come back to this point later.

\section{Results and Discussions for the 2D system}

\subsection{Open shell fillings}
\begin{figure}[htb]
\begin{center}
\scalebox{0.8}{
\includegraphics[width=10cm,clip]{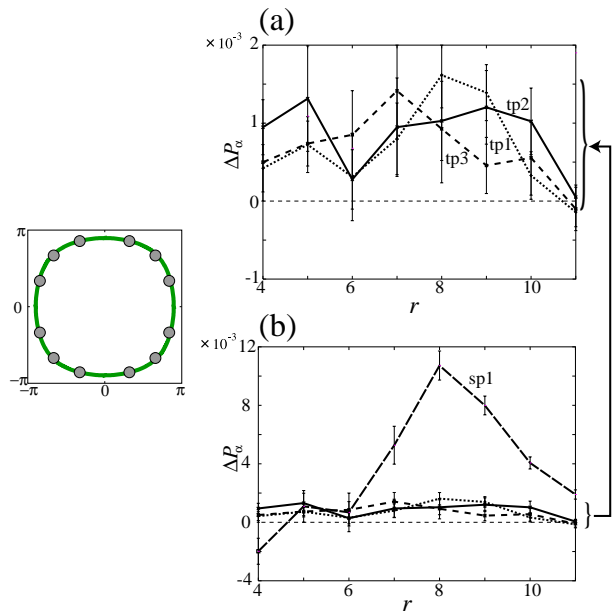}}
\caption{
The QMC result of $\Delta P_\alpha$
for a $N=12\times 12$ 2D 
system with an open shell filling close to $n\sim 4/3$. 
(a) focuses only on the triplet channels, while (b) shows 
all the channels considered.
Parameter values are ($N_e$, $n$, $t'$)=(190, 1.32, 0.464).
\label{fig7}}
\end{center}
\end{figure}

\begin{figure}[htb]
\begin{center}
\scalebox{0.8}{
\includegraphics[width=10cm,clip]{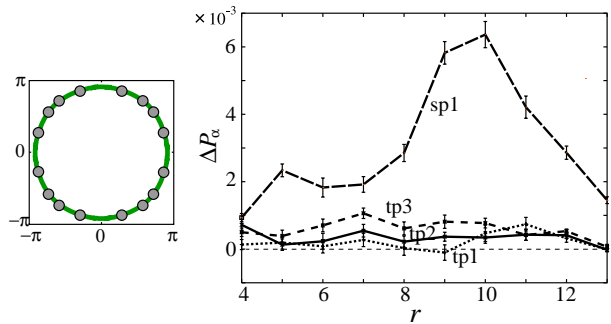}}
\caption{
A plot similar to Fig.\protect\ref{fig7} but for a $N=14\times 14$
 system. Parameter values are ($N_e$, $n$, $t'$)=(258, 1.32, 0.406).
\label{fig8}}
\end{center}
\end{figure}

Let us now move on to the 2D system with $n\simeq 4/3$.
We first look into the case with open shell fillings.
In Fig.\ref{fig7}(a), we focus on the interaction vertices for 
triplet pairings, where we find positive interaction vertices for
all three channels. 
However, if we compare singlet and triplet 
channels in Fig.\ref{fig7}(b), $\Delta P_{\rm sp1}>\Delta P_{\rm tp}$
holds for all three triplet channels. A similar result is 
obtained also for a $14\times 14$ system as seen in Fig.\ref{fig8}.
Since the pairing forms are not optimized  
and there is a possibility of stronger triplet pairings 
between more separated sites than we have considered,
we cannot conclude that singlet dominates over triplet. 
However, considering the fact that this is a result for an open shell filling,
we may say that $d_{x^2-y^2}$-wave is 
still a strong candidate for the most dominant pairing channel 
even though the band filling is considerably away from half filling 
and $t'$ is large.

This result provides an example of 
singlet pairing vertex being larger than 
that of the triplet pairings even for an open shell filling, 
which conversely supports our argument in section \ref{4a} 
that $\Delta P_f >\Delta P_d$ in the quasi-1D system 
may not simply be an artifact of the open shell filling, 
and that $f$-wave may be competitive against $d$-wave 
even in the on-site Hubbard model due to 
effects beyond RPA-type theories.

\begin{figure}[htb]
\begin{center}
\scalebox{0.8}{
\includegraphics[width=10cm,clip]{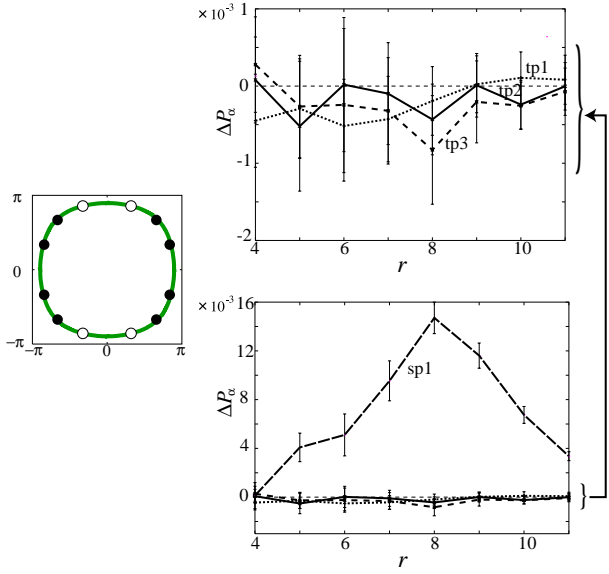}}
\caption{
A plot similar to Fig.\protect\ref{fig7}(b) but for a closed shell filling.
Parameter values are ($N_e$, $n$, $t'$, $t_y$)=(194, 1.35, 0.465, 0.999).
\label{fig9}}
\end{center}
\end{figure}

\begin{figure}[htb]
\begin{center}
\scalebox{0.8}{
\includegraphics[width=10cm,clip]{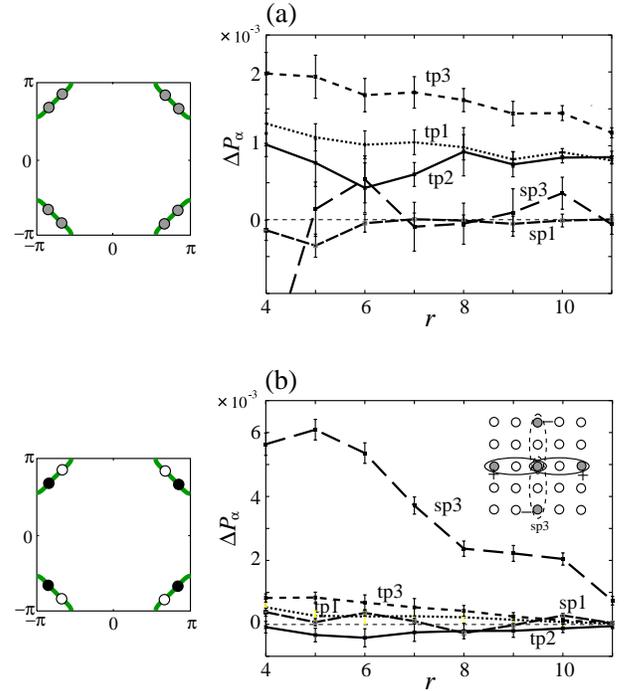}}
\caption{
The QMC result of $\Delta P_\alpha$
for a $12\times 12$ 2D system near full filling 
for an open shell (a) and for a closed shell (b).
Parameter values are ($N_e$, $n$, $t'$)=(254, 1.76, 0.400).
$t_y=0.999$ is taken in (b). The sp3 pairing is shown in the 
right inset of (b).
\label{fig10}}
\end{center}
\end{figure}

\begin{figure}[htb]
\begin{center}
\scalebox{0.8}{
\includegraphics[width=10cm,clip]{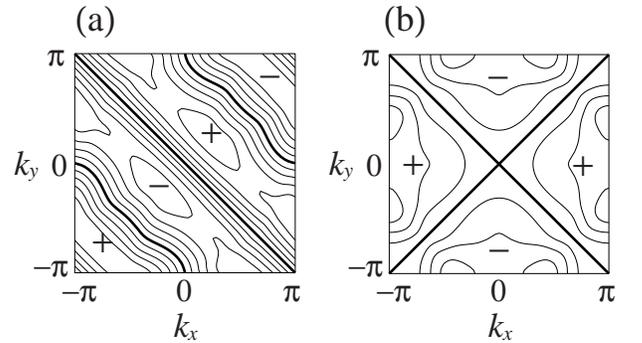}}
\caption{
Contour plots of the 
gap functions for (a) the triplet and (b)the singlet channels 
obtained by FLEX + Eliashberg equation for the 
2D Hubbard model with $t'=0.4$, $n=4/3$, $U=4$, and $T$(temperature in
 units of $t$)=0.005. The thick lines represent the nodal lines of the 
gap function.
\label{fig11}}
\end{center}
\end{figure}

\subsection{Closed shell filling}
In Fig.\ref{fig9}, we present results for a closed shell filling.
The interaction vertex for sp1 pairing 
becomes larger than for the open shell filling (Fig.\ref{fig7}).
The vertices for triplet pairings this time take negative 
values for most of the distances.
This is in contrast with 
the result for the quasi 1D system in Fig.\ref{fig5}, where $\Delta_f$
remains positive even for a closed shell filling.
This comparison further enforces our argument that $f$-wave pairing 
is indeed a favorable pairing in the quasi-1D system.

\subsection{Near full filling}
To make sure that we have been on the right track,
we present here QMC results for a nearly full filled (dilute hole
density) case, where the outcome can be safely predicted.
Namely in this case, 
spin triplet pairing is expected to dominate strongly over singlet 
since ferromagnetic spin fluctuations arise for 
$t'\sim 0.5$ and $n \sim 2$ (or equivalently $t'\sim -0.5$ and $n\sim
0$) due to the large density of states at the Fermi level.
This can be considered as well established because a number of approaches, 
including a FLEX study\cite{AKA,comment2}, 
perturbational studies\cite{Chubukov,Fukazawa}, a $T$-matrix approximation
\cite{Hlubina}, and a renormalization group study\cite{Honerkamp}, 
reach (at least qualitatively) the same conclusion in this parameter regime.

Since pairs may be formed at large distances for dilute career density, 
here we calculate the pairing interaction vertex 
also for the singlet sp3 channel shown in the inset of 
Fig.\ref{fig10}(b).
For an open shell filling, the interaction vertices are in fact positive for 
all three triplet channels, while they are small or negative 
for the singlet sp1 and sp3 channels  
as seen in Fig.\ref{fig10}(a). 
By contrast, for a closed shell filling, 
although the interaction vertices for 
tp1 and tp3 remains positive, they are much smaller than that 
for the singlet sp3 channel. This result contradicts with the 
established result, again supporting our
argument that the result $\Delta P_d > \Delta P_f$ 
for a closed shell filling in the quasi 1D system (Fig.\ref{fig6}) 
may not be taken directly.

\subsection{Comparison with FLEX}
\label{VD}
To close this section, we present FLEX\cite{Bickers} calculation results 
for the 2D on-site Hubbard model with $n=4/3$, 
$t'=0.4$ and $U=4$, and compare them with the QMC results given above. 
The formulation is as follows. First, the renormalized Green's function
is obtained within the FLEX method, which is kind of a self-consistent 
RPA. Secondly, the pairing interaction mediated 
by spin fluctuations is calculated within the RPA form.  
Finally, the FLEX Green's function and the pairing interaction
are substituted for the linearized Eliashberg equation, 
which is solved using the power method to give the 
largest eigenvalue and the corresponding eigenfunction 
(the gap function).

The gap functions having the largest eigenvalues 
are shown in Fig.\ref{fig11} 
for (a) the triplet and (b) the singlet pairing channels.
The singlet gap function has a form close to $\cos(k_x)-\cos(k_y)$,
which corresponds to sp1 pairing in the QMC study, while the form of the 
triplet gap function is close to $\sin(k_x+k_y)$, which corresponds to
tp2. Although the eigenvalues are small (less than 0.3 for $T\geq
0.005$) for both singlet and triplet,
the ratio between the singlet and triplet eigenvalues 
turns out to be 
$\lambda_{\rm singlet}/\lambda_{\rm triplet}\simeq 5$ at $T=0.005$, which
is at least qualitatively consistent with the QMC results.

\section{Summary}

To summarize, we have studied the competition among 
various pairing symmetries in the quasi-1D and the 2D 
on-site Hubbard models  
by calculating the pairing interaction vertices using the 
ground state QMC technique.
For the quasi-1D system, $f$-wave pairing not only 
dominates over $p$-wave in agreement with 
the spin-charge fluctuation theory, but also looks 
competitive against $d$-wave pairing. 
$f$-wave being competitive against $d$-wave 
in the on-site Hubbard model is rather unexpected from 
the spin-charge fluctuation theory because 
$2k_F$ charge fluctuation is not enhanced in the on-site 
Hubbard model.\cite{Hirsch} If $f$-wave is competitive 
against $d$-wave {\it even} in the absence of strong 
charge fluctuations, it is likely that $f$-wave dominates 
over $d$-wave in the actual (TMTSF)$_2$X, 
where $2k_F$ CDW actually coexists with SDW.\cite{PR,Kagoshima}

For the 2D system, the calculation results (at least for the 
open shell filling) support presence of attractive interaction
 in the triplet pairing channel. 
However, singlet $d_{x^2-y^2}$ pairing dominates 
over all the triplet pairings considered in the present study 
{\it at the band filling $n\simeq 4/3$}, which corresponds to Sr$_2$RuO$_4$,
in agreement with the FLEX calculation. 
Although we cannot completely 
rule out the possibility of other (longer ranged) triplet 
pairings dominating over singlet, or a possibility of 
triplet dominating over singlet for larger system sizes 
and/or larger values of $U$, 
the present results do suggest that it is worth considering 
effects beyond the single band Hubbard model, including the 
contributions from the $\alpha-\beta$ bands,\cite{KO,SK,KOAA,Takimoto,Eremin} 
in order to fully understand the occurrence of triplet superconductivity in 
Sr$_2$RuO$_4$.

%
Numerical calculation has been performed
at the facilities of the Supercomputer Center,
Institute for Solid State Physics,
University of Tokyo, and at the Computer Center, 
University of Tokyo.

%



\begin{thebibliography}{99}
\bibitem{commentd-f} At very short distances not shown in the figures,
$\Delta_d$ is larger than $\Delta_f$. However, 	
this only reflects the fact that spins tend to be aligned 
in the 2$k_F$ SDW configuration, where the 
{\it spin density} 
correlation $\langle n_{i\uparrow}n_{i+2\downarrow}\rangle$ is large.
It is the long range part of the pairing correlation 
that provides the true measure 
for the pairing interaction, so we focus only on this part.
\bibitem{comment2} Although a finite $T_c$ in the triplet channel 
is not obtained in the FLEX calculation, the eigenvalue of the Eliashberg
equation is larger in the triplet channel than in the singlet channel.
\end{thebibliography}
\end{document}